\documentclass{article}
\usepackage{amsfonts,amsmath,amssymb,mathrsfs,bigints}
\usepackage[english]{babel}

\def\g{{\gamma}}
\newtheorem{prop}{Proposition}

\begin{document}

%%%%%%%%%%%% TITLE %%%%%%%%%%%%%%

\title{Poisson structures for two nonholonomic systems with partially reduced symmetries}
\author{ A V Tsiganov \\
\it\small
St.Petersburg State University, St.Petersburg, Russia\\
\it\small e--mail:  andrey.tsiganov@gmail.com}

%\subjclass{Primary: 34D20; Secondary: 70E40, 37J35.}
%\keywords{integrable nonholonomic systems, Poisson geometry}
%\email{andrey.tsiganov@gmail.com}
%\thanks{This work was partially supported by RFBR grant 13-01-00061.}

\date{}
\maketitle

\begin{abstract}
We consider nonholonomic systems which symmetry groups consist of two subgroups one of which represents rotations about the axis of symmetry. After nonholonomic reduction by another subgroup the corresponding vector fields on partially reduced phase space are linear combinations of the Hamiltonian and symmetry vector fields. The reduction of the Poisson bivectors associated with the Hamiltonian vector fields to canonical form is discussed.
 \end{abstract}
 \maketitle

%%%%%%%%%%%%%%%%%%%%%%%%%%%%%%%%%%%%%%%%%
\section{Introduction}

 In 1887-1911 S.A. Chaplygin proposed a theory of reducing multiplier for investigation of nonholonomically constrained mechanical systems with an invariant measure \cite{ch48,ch11}. The first part of this theory is a nonholonomic reduction of the phase space $\mathcal M$ by a symmetry group $G$, which is based on the most famous and established method for finding exact solutions of differential equations, which is called the classical symmetries method or group analysis, which originated in 1881 by S. Lie. The second part states that nonholonomic system on the reduced phase space $\mathcal M/G$ became Hamiltonian after a suitable reparameterization of time, a process now referred to as Chaplygin Hamiltonization \cite{bm01,ekm04,jov09,hoc09,bl11,st89}.

It is known that sometimes symmetry group $G$ consists of various subgroups $G=G_1\times G_2$ associated either with the translations and the different kinds of rotations \cite{cus98,herm95,zen95}, or with external and internal symmetries, see \cite{bates93,ekm04,cus10} and reference within. The existence of some subgroups allows us to make a partial nonholonomic reduction $\mathcal M/G_1$ by one of the subgroups \cite{cus10}, then to reduce the corresponding Poisson bracket to canonical one and only then to finish reduction by remaining subgroup $G_2$. Thus, we want to construct an appropriate Poisson map which identifies the given dynamical system on the partially reduced phase space with a dynamical system on some well-known manifold equipped with the canonical Poisson brackets. This reduction to the canonical Poisson brackets may be useful for the comparison of the different nonholonomic system to each other and to the application of the well-studied methods of Hamiltonian mechanics to the nonholonomic systems. For instance, in \cite{ts12b,ts12c} we used such reduction in order to prove the equivalence of the Chaplygin ball problem with the nonholonomic Veselova problem.

Below we consider the motion of the body of revolution on a plane and motion of the homogeneous ball on the surface of revolution. For both these systems initial phase space is $\mathcal M=TE(3)$, where $E(3)$ is the Euclidean group of all rigid motions. The symmetry groups $G=G_1\times G_2$ consist of different subgroups $G_1=E(2)$ or $G_1=SO(3)$, and the common subgroup $G_2=SO(2)$ represents rotations about the corresponding axis of revolution. The main aim of this paper is to identify the  common level surfaces of integrals of motion with lagrangian foliation with respect to some Poisson bivector. Later we will try to  identify the  common level surfaces of integrals of motion with bi-lagrangian foliation with respect to some Poisson pencil and then with some subvariety of the Jacoby variety associated with some algebraic curve related with this pencil.

\subsection{Vector fields on partially reduced phase space}
Let us consider some smooth manifold $\mathcal M$ with coordinates $x_1,\ldots x_m$ and a dynamical system defined by the following equations of motion
\begin{equation}\label{d-eq}
\dot{x}_i=X_i\,,\qquad i=1,\ldots,m.
\end{equation}
This system of ODE's defines a vector field
\begin{equation}\label{d-vp}
X=\sum_{i=1}^m X_i\dfrac{\partial}{\partial x_i}\,,
\end{equation}
which is a linear operator on a space of smooth functions on $\mathcal M$ that encodes the infinitesimal evolution of any quantity
\[
\dot{F}=X(F)=\sum X_i \dfrac{\partial F}{\partial x_i}
\]
along with the solutions of the system of equations (\ref{d-eq}).

In Hamiltonian mechanics any Hamilton function $H$ on $\mathcal M$ generates vector field $X$ describing the dynamical system
\begin{equation}\label{ham-vp}
X= P\mathrm dH\,.
\end{equation}
Here $\mathrm dH$ is a differential of $H$ and $P$ is a Poisson bivector on the phase space $\mathcal M$.

For a lot of nonholonomic dynamical systems vector fields on the reduced phase space are created using the Hamilton function $H$ and the reducing multiplier $g$
 \begin{equation}\label{cham-vp}
 X=gP\mathrm dH\,.
 \end{equation}
 This vector field $X$ (\ref{cham-vp}) is the so-called conformally Hamiltonian vector field, see examples of such fields in \cite{bm01,bm02,ts12a,ts12b}.

In this note we discuss the rolling motion of a heavy rigid body of revolution on a rough horizontal plane and the motion of a homogeneous ball on the surface of revolution \cite{ch48,rou55}. For both of these nonholonomic systems vector fields $X$ on the partially reduced six-dimensional phase space are linear combinations of the Hamiltonian field $P\mathrm dH$ and the symmetry vector field $X_S$ associated with the rotations about the axis of revolution:
\begin{equation}\label{d-gen}
 X=P\mathrm dH+\eta X_S\,.
 \end{equation}
According to the Noether's theorem if the symmetry field $X_S$ is related with the integral of motion $J$, then we can rewrite (\ref{d-gen}) in the following form
\begin{equation}\label{vf-dec}
X=P\mathrm dH+\nu P\mathrm dJ\,.
\end{equation}
Below we present the corresponding Poisson bivectors $P$, rank\,$P=4$, which are missed in the existing works on these nonholonomic systems. Moreover, we show that this bivector $P$ is a deformation of the canonical Poisson bivector on the Lie algebra $e^*(3)$.

According to \cite{ch48}  projection of $X$ (\ref{vf-dec}) on  the four dimensional submanifold $\mathcal M/G$  is the Hamiltonian vector field. This Chaplygin idea of reduction has generated considerable interest and has been applied to various nonholonomic systems, see \cite{bates93,bm02,hoc09,koi92,bl11,wor09,zen95} and the references within.
However, projection of $P$ on the invariant submanifold $\mathcal M/G$ is usually a highly degenerate Poisson bivector. Thus we could have some problems with the definition of the symplectic leaves and the Casimir functions, which have to be solved before creating suitable Hamiltonian vector fields on these symplectic leaves.

\subsection{ Euler-Jacobi theorem and rank-two Poisson structures}
Let us consider the following system of differential equations
\[
\dfrac{\mathrm dx_1}{X_1}=\dfrac{\mathrm dx_2}{X_2}=\cdots=\dfrac{\mathrm dx_k}{X_k}\,,
\]
 where $X_i$ are the functions on variables $x_1,\ldots,x_k$. If we know the Jacobi multiplier $\mu $ defined by
\begin{equation}\label{y-mera}
\dfrac{\partial \mu X_1}{\partial x_1}+\dfrac{\partial \mu X_2}{\partial x_2}+\cdots+\dfrac{\partial \mu X_k}{\partial x_k}=0\,,
\end{equation}
and know $k-2$ independent first integrals $H_1,\ldots, H_{k-2}$, we can integrate this system by quadratures.

Namely, according to the Euler-Jacobi theorem, in this case we can introduce new variables $y_1,\ldots,y_k$ by rule
 \[
 y_1=x_1,\qquad y_2=x_2,\qquad y_3=H_1,\quad\ldots\quad,\, y_k=H_{k-2}\,.
 \]
In $y$-variables the initial system of equations has the following form
\begin{equation}\label{eqy-in}
\dfrac{\mathrm dy_1}{Y_1}=\dfrac{\mathrm dy_2}{Y_2}=\dfrac{\mathrm d y_3}{Y_3}=\cdots=\dfrac{\mathrm dy_k}{Y_k}\,,\qquad\mbox{where}\qquad Y_i=0\,,\quad i\geq 3\,,
\end{equation}
or
\[Y_2\mathrm dy_1-Y_1\mathrm dy_2=0\,.\]
Using the definition (\ref{y-mera}) of the Jacobi multiplier $\mu$
\[
\dfrac{\partial \mu Y_1}{\partial y_1}+\dfrac{\partial \mu Y_2}{\partial y_2}=0\,,
\]
we can prove that $\mu (Y_2\mathrm dy_1-Y_1\mathrm d y_2)$ is the total differential. So, there is one more independent first integral
\[
H_{k-1}=\int \mu (Y_2\mathrm dy_1-Y_1\mathrm d y_2)\,.
\]
It is well-known that any $k$-dimensional dynamical system with the independent $k-1$ first integrals and the Jacobi multiplier $\mu$ may be rewritten in the Hamiltonian form
\begin{equation}\label{eqy-ham}
Y=P^{(y)}\mathrm d H\,,
\end{equation}
according to the Vall\'{e}e-Poussin theorem on functional determinants \cite{vp38}. Here $P^{(y)}$ is a rank-two Poisson bivector and $H$ is a function on $H_1,\ldots,H_{k-1}$, see details in \cite{bm99} .

For instance, we can put
\begin{equation}\label{y-poi}
P^{(y)}_{12}=\mu ^{-1}\,,\qquad P^{(y)}_{13}=Y_1\,,\qquad P^{(y)}_{23}=Y_2\,,
\end{equation}
so that
\[
\{y_1,y_2\}=\mu^{-1}\,,\qquad \dot{y}_1=\{y_1,H\}=Y_1\,,\qquad \dot{y}_2=\{y_2,H\}=Y_2\,,
\]
and
\[\dot{y_i}=0\,,\quad i\geq3\,.\]
Other brackets are equal to zero. The corresponding rank-two bivector $P^{(y)}$ is the Poisson bivector iff
\[
\dfrac{\partial \mu Y_1}{\partial y_1}+\dfrac{\partial \mu Y_2}{\partial y_2}=-{\mu^2Y_1^2}\,\dfrac{\partial }{\partial y_3} \dfrac{Y_2}{Y_1}\,.
\]
If we choose variables $y_1,\ldots,y_k$ so that the ratio $Y_2/Y_1$ is independent on $y_3$, then the Hamiltonization of dynamical equations (\ref{eqy-ham}) is equivalent to the existence of invariant measure because
\[
[P^{(y)},P^{(y)}]=0\,,\qquad\Leftrightarrow\qquad \sum_{i=1}^k \dfrac{\partial \mu X_i}{\partial y_i}=0\,.
\]
Here $[.,.]$ is a Schouten bracket. This construction of rank-two Poisson bivectors for the nonholonomic St\"{u}bler model is discussed in \cite{ts13a}.

Such rank-two Poisson structures are well-studied \cite{bm02,fass05,herm95,mosh87,ram04} and, therefore, we will only consider rank-four Poisson structures below.

\section{Rolling motion without sliding}

In \cite{app96,app00, ch48, bm02, kor00, nf67, not09, rou55} there are variants of the derivation of the equations of rolling motion of a heavy rigid body, which touches a surface at one point. For brevity we reproduce only the necessary facts about the motion of bodies of revolution on the plane and facts about the motion of the homogeneous ball on the surface of revolution. In both cases the equations of motion on six-dimensional phase space have a rotational symmetry about the axis of revolution.

\subsection{Body of revolution on a plane}
The moving body is subject to two kinds of constraints: a holonomic constraint of moving over of a horizontal plane and no slip nonholonomic constraint associated with the zero velocity at the point of contact
\begin{equation}\label{rel-ch}
v+\omega\times r=0.
\end{equation}
Here $\omega$ and $v$ are the angular velocity and velocity of the center of mass of the body, $r$ is the vector joining the center of mass with the contact point and $\times$ means the vector product in $\mathbb R^3$. All the vectors are expressed in the so-called body frame, which is firmly attached to the body, its origin is located at the center of mass of the body, and its axes coincide with the principal inertia axes of the body.

In the body frame the angular momentum $M$ of the body with respect to the contact point is equal to
\begin{equation}\label{om-m}
M=\mathbf{J} \omega\,,\qquad \mathbf{J}=\mathbf{I}+mr^2\mathbf E-mr\otimes r.
\end{equation}
Here $\mathbf E$ is a unit matrix, $m$ is a mass and $\mathbf I = \mathrm{diag}(I_1, I_2, I_3 )$ is an inertia tensor of the rolling body.

After elimination of the Lagrange multiplier and reduction by $E(2)$ one gets the following equations of motion
\begin{equation}\label{rotb-eqm}
\dot M=M\times \omega+m\dot{r}\times(\omega\times r)+M_F\,,\qquad
\dot \g=\g\times \omega\,,
\end{equation}
where $\g$ is the unit vector orthogonal to the plane at the contact point and $M_F$ is the moment of external forces with respect to the contact point depending on $\g$ only. Vector $r$ is expressed as a function of normal vector $\g$ using the following relation
\[
\gamma=-\frac{\mathrm{grad}\, \mathfrak f}{|\mathrm{grad}\,  \mathfrak f|}\,,
\]
 that defines the Gauss transformation, if $\mathfrak f(r)=0$ is an equation of the body surface.

Intermediate five dimensional reduced space $\mathcal M/G_1$ may be identified with the submanifold $||\g||=1$ in the six-dimensional space $(\g,M)\in\mathbb R^3\times\mathbb R^3$. Similar to the Lagrange top  we will study the vector field $X$ defined by equations (\ref{rotb-eqm}) on the six-dimensional manifold with local coordinates $x=(\g_1,\g_2,\g_3,M_1,M_2,M_3)$. Restriction $||\g||=1$ we will consider as an integral of motion for these equations in order to have some additional freedom for the search of the Poisson bivectors, see  Section 2.3.

 If the moment of external forces is described by potential $U$ depending only on $\g$ we have
 \[M_F=\gamma\times\frac{\partial U}{\partial \gamma}\,.\]
In this case vector field $X$ (\ref{rotb-eqm}) possesses two integrals of motion
\begin{equation}\label{rotb-ham}
H=\frac12 (M,\omega)+U(\gamma)\,,\qquad C=(\gamma,\gamma)=1\,.
\end{equation}
In some cases vector field $X$ has an invariant measure and additional integrals of motion. In the next paragraph we consider one of these particular cases.

If we want to study a rigid body of revolution we have to impose some restrictions
\begin{equation}\label{symm-eq}
I_1=I_2\neq I_3\,,\qquad r_1=f_1(\gamma_3)\gamma_1\,,\quad r_2=f_1(\gamma_3)\gamma_2\,,\quad r_3=f_2(\gamma_3)\,.
\end{equation}
on the tensor of inertia $\mathbf I=(I_1,I_2,I_3)$ and on the equation of the body surface $\mathfrak f(r)=0$.

It means that the surface of the body and its central ellipsoid of inertia are coaxial surfaces of revolution.
Because $(\dot{r},\g)=0$ functions $f_{1,2}(\g_3)$ in (\ref{symm-eq}) satisfy the equation
\[
\dfrac{\mathrm df_2(\g_3)}{\mathrm d\g_3}=f_1(\g_3)-\dfrac{1-\g_3^2}{\g_3}\,\dfrac{\mathrm df_1(\g_3)}{\mathrm d\g_3}\,,
\]
that defines a meridional section. Below we do not use this equation at all.

We also assume that $U$ is an arbitrary function of $\g_3$, i.e.  that it depends only on the slope of the revolution axis of the body to the vertical \cite{ch48}. In particular, it means that in the case of gravity field the center of mass must be situated on the axis of revolution.

Under the above conditions the symmetry group $G$ consist of two independent parts \cite{cus98}. The first part of the symmetry group $G_1=E(2)$ is generated by translations of the horizontal plane and rotations about the vertical axis while the second part $G_2=SO(2)$ is generated by rotations about the axis of revolution. On the six-dimensional phase space vector field $X$ (\ref{rotb-eqm}) has a symmetry field associated with the second part of the symmetry group
\begin{equation}\label{vf-symm}
X_S=\gamma_1\dfrac{\partial }{\partial \gamma_2}-\gamma_2\frac{\partial}{\partial \gamma_1}+M_1\dfrac{\partial }{\partial M_2}-M_2\frac{\partial}{\partial M_1}
\end{equation}
and an invariant measure
\begin{equation}\label{rotb-mu}
\rho=g^{-1}(\gamma)\,d\gamma\, dM\,,\qquad g(\gamma)={\sqrt{I_1I_3+m(r,\mathbf I r)}}\,.
\end{equation}
According to \cite{app96,app00,ch48,bm02} there are also two additional integrals of motion, which are linear in momenta functions
\begin{equation}\label{rotb-addint}
J_{k}=v_1^{(k)}(\g_3)(\g_1M_1+\g_2M_2)+v_2^{(k)}(\g_3)M_3\,,\qquad k=1,2.
\end{equation}
In generic case coefficients $v_1^{(k)}$ and $v_2^{(k)}$ are the real analytic non algebraic functions on $\g_3$, which satisfy to the following system of equations:
 \begin{equation}\label{rotb-eqv}
 \begin{array}{l}
\dfrac{g^2(v'^{(k)}_2+v^{(k)}_1)}{m}=mf_1f_2\bigl((1-\g_3^2)f_1+\g_3f_2\bigr)(v^{(k)}_2f_1-v^{(k)}_1f_2)\\
\qquad+I_1f_1\Bigl((1-\g_3^2)(v^{(k)}_1f'_2-v^{(k)}_2f'_1)-\g_3v^{(k)}_1f_2+\bigl(v^{(k)}_1(\g_3^2-1)+\g_3v^{(k)}_2\bigr)f_1\Bigr)\,,\\
\\
\dfrac{g^2v'^{(k)}_1}{m}=mf_1^2\bigl((1-\g_3^2)f_1+\g_3f_2\bigr)(v^{(k)}_2f_1-v^{(k)}_1f_2)\\
\\
\qquad+I_3\Bigl(f_1^2v^{(k)}_2-(v^{(k)}_1f'_2-v^{(k)}_2f'_1)f_2\Bigr)\,.
\end{array}
\end{equation}
Here and everywhere below we omit dependence of functions on $\g_3$, i.e.
 \[f(\g_3)=f\,,\qquad f'=\frac{df(\g_3)}{\mathrm d\g_3}\,.\]

 For example, let us consider a rolling disk of radius $R$ with the center of mass displaced on $a$ along the axis of dynamical symmetry. In this case
\[
f_1=\dfrac{R}{\sqrt{1-\g_3^2}}\,,\qquad f_2=a\,.
\]
 At $a=0$ functions $v_{1,2}^{(k)}$ are equal to
\begin{equation}\label{disk-int}
v_1^{(k)}=L^{(k)}\left(b_-,\g_3\right)\,,\qquad v_2^{(k)}=c\left(L^{(k)}(b_+,\g_3)-\g_3L^{(k)}(b_-,\g_3)\right)\,,\qquad k=1,2,
\end{equation}
where
\[b_{\mp}=\frac{\sqrt{g^2-4mI_3R^2}}{2gI_1^{1/2}}\mp \frac12\,,\qquad c=-\frac{I_1^{1/2}\sqrt{g^2-4mI_3R^2}+gI_1}{2mgR^2}
\]
and $L^{(1,2)}(b_k,\g_3)$ are the Legendre functions of the first and second kind, respectively. However, at $a\neq 0$ we have only implicit definitions of the additional integrals of motion (\ref{rotb-addint}).

\subsection{Homogeneous ball on a surface of revolution}
Let us consider the homogeneous ball with mass $m$, radius $R$ and tensor of inertia $\mathbf I = \mu\mathbf E$, where $\mathbf E$ is a unit matrix. We are going to study the case when the ball rolls without sliding on the surface having only one point in common during all motion.

The surface is defined by equation
\[\mathfrak f(r)=0\,\]
and the rolling ball is subject to two kinds of constraints: a holonomic constraint of moving over the surface and no slip nonholonomic constraint associated with the zero velocity at the point of contact
\[v+\omega\times a=0\,,\]
where we denote the velocity of the ball's center by $v$, the angular velocity of the ball by $\omega$ and vector joining the ball's center with the contact point by $a$.

If $N$ is the reaction force at the contact point, $F$ and $M_F$ are the external force and its moment with respect to the contact point, then the conservation principles of linear and angular momentum read as
\[
m\dot{v}=N+F\,,\qquad \dot{\mathbf I \omega}=a\times N+M_F\,.
\]
After substituting $a=-R\g$, where $\g$ is a normal to the surface, and eliminating the reaction force $N$ one gets the following equations of motion
\begin{equation}\label{ball-eqm}
\dot{M}=d\dot{\g}\times (\omega\times\g)+M_F\,,\qquad \dot{r}+R\dot{\g}=\omega\times R\g\,,
\end{equation}
which define vector field $X$ on the six-dimensional phase space with local coordinates $x=(\g_1,\g_2,\g_3,M_1,M_2,M_3)$. Here $d=mR^2$ and vectors $\omega$ and $r$ are functions on $x$ defined by the following relations
\[
M=\mu\omega+d\g\times(\omega\times\g)\,,\qquad\g=\dfrac{\nabla \mathfrak f(r)}{|\nabla \mathfrak f(r)|}\,.
\]
If external force $F$ is a potential force associated with potential $U=U(r+R\g)$, then
\[M_F=R\g\times\dfrac{\partial U}{\partial r'}\,,\qquad r'=r+R\g\,,
\]
and equations (\ref{ball-eqm}) possess first integrals
\[
H=\frac12(M,\omega)+U(r')\,,\qquad C=(\g,\g)=1\,.
\]

Below we consider the homogeneous ball rolling on the inner side of the surface of revolution obtained by rotation of a smooth curve about a vertical axis. Following \cite{bmk02,not09, rou55}, we use parameterization of the surface on which the ball's center of mass is moving, so that
\[
r_1=(f-R)\g_1\,,\qquad r_2=(f-R)\g_2\,,\qquad
r_3=\int\left(f-(1-\g_3^2)\g_3^{-1}f'\right)\mathrm d\g_3-R\g_3\,,
\]
where $f=f(\g_3)$ is a sufficiently smooth function. It means that the curvature of the meridian of surface is smaller than $1/R$.
 Moreover, for brevity, we consider free motion at $U=0$, because free motion and potential motion are usually associated with the same Poisson brackets.

Apparently Routh was the first to explore this problem. In \cite{rou55} he described the family of stationary periodic motions, obtained a necessary condition for stability of these motions and also noticed that the integration of the equations of motion may be reduced to an integration of a system of two linear differential equations with variable coefficients and considered several cases when the equations of motion can be solved by quadratures.

In this case the symmetry group consists of two independent parts $G=SO(3)\times SO(2)$ \cite{fass05,zen95}. The first part is generated by rotations of the ball about its center while the second part is generated by rotations about the vertical axis of revolution. In contrast with the body of revolution on a plane all the vectors are expressed in the space frame. In some sense change of frames allows us to "exchange" inner and external symmetry subgroups according to  \cite{cus10}.

As above we will extend the five dimensional reduced space $\mathcal M/G_1$ to the six-dimensional space $(\g,M)\in\mathbb R^3\times\mathbb R^3$ and will consider restriction $||\g||=1$ as an integral of motion.
On the six-dimensional phase space vector field $X$ (\ref{rotb-eqm}) has a symmetry field (\ref{vf-symm})
\[
X_S=\gamma_1\dfrac{\partial }{\partial \gamma_2}-\gamma_2\frac{\partial}{\partial \gamma_1}+M_1\dfrac{\partial }{\partial M_2}-M_2\frac{\partial}{\partial M_1}\,.
\]
and  invariant measure
\begin{equation}\label{ball-invm}
\rho=g^{-1}(\g)\,d\g dM\,,\qquad g^{-1}(\g)=f^3\left(f-(1-\g_3^2)\g_3^{-1}f'\right)\,,
\end{equation}
that means that vector field $X$ may be integrated by quadratures.

As above, there are  linear in momentum integrals of motion
\begin{equation}\label{ball-addint}
J_{k}=v_1^{(k)}(\g_1M_1+\g_2M_2)+v_2^{(k)}M_3\,,\qquad k=1,2.
\end{equation}
Here functions $v_{1,2}(\g_3)$ are the solutions for the following system of equations
\begin{equation}\label{ball-eqv}
v'_1=\frac{f'\bigl(v_1(\mu+ \g_3^2d)-\g_3 v_2d\bigr)}{f(\mu+d)}\,,\qquad
v'_2=v_1-\frac{f'(1-\g_3^2)\bigl(v_1(\mu+ \g_3^2d)- \g_3 v_2d\bigr)}{\g_3f(\mu+d)}\,.
\end{equation}
Relations of these integrals with symmetries is discussed in \cite{bmk02,fass05,rou55,zen95}.

For example, let us suppose the ball's center is moving on the ellipsoid of revolution with the principal semi-axes $b^{1/2}_{1,3}$. In this case
\[
f(\g_3)\equiv f=-\frac{b_1}{b_1(1-\g_3^2)+b_3 \g_3^2}
\]
and equations (\ref{ball-eqv}) have the following two independent solutions
\begin{eqnarray}\label{ball-ellv}
v_1^{(1,2)}&=&b_1^{-1}\g_3^{3/2}f\,L^{(1,2)}\left(\nu-\frac{1}{2},\frac{3}{2},\frac{\sqrt{b_1}}{f}\right)\,,\\
 v_2^{(1,2)}&=&\g_3^{1/2}f\left(\frac{\g_3^2-1}{b_1}+\frac{1}{(b_1-f^2)(\nu+1)}\right)
 L^{(1,2)}\left(\nu-\frac{1}{2},\frac{3}{2},-\frac{\sqrt{b_1}}{f}\right)\nonumber\\
 &+&\frac{\g_3^{1/2}f^2}{\sqrt{b_1}(b_1-f^2)(\nu+1)}L^{(1,2)}\left(\nu+\frac{1}{2},\frac{3}{2},-\frac{\sqrt{b_1}}{f}\right)\,,\nonumber
\end{eqnarray}
where $L^{(1,2)}$ are the associated Legendre functions of the first and second kind and
\[
\nu=\sqrt{\frac{\mu}{\mu+d}}\,.
\]
According to \cite{bmk02} there is also the second order in momenta algebraic integral of motion
\[
H_2=\frac{b_1^2\mu(\g,M)^2}{\mu+d}+(b_1-b_3)\left(\frac{f\mu\bigl(\g_3(\g,M)-M_3\bigr)}{\mu+d} \right)^2
\]
and equations of motion may be solved in the elementary functions.

\subsection{The Poisson structure for integrable nonholonomic systems}

In the Chaplygin theory of reducing multiplier we have a constructive algorithm for calculation of the Poisson structures on the completely reduced submanifold $\mathcal M/G$ only. In order to get the Poisson structure on the partially reduced six-dimensional phase space $\mathcal M/G_1$ with the coordinates $x=\g,M$ we have to apply other methods.

 In this paper we use the brute force approach proposed in \cite{ts11}. Namely, six equations of motion (\ref{rotb-eqm}) and (\ref{ball-eqm}) possess four integrals of motion
\[H_1=H\,,\qquad H_2=C\,,\qquad H_3=J_1\,,\qquad H_4=J_2\]
and the invariant measure (\ref{rotb-mu},\ref{ball-invm}) and, therefore, they are integrable by quadratures according to the Euler-Jacobi theorem.

If we  identify the common level surfaces of integrals $H_1\,,\ldots,H_4$ with the Lagrangian foliation of symplectic leaves of some unknown rank-four  Poisson bivector $P$, then  the following equations have to be fair
 \begin{equation}\label{geom-eq}
[P,P]=0\,,\qquad \{H_k,H_m\}=\sum_{i,j=1}^6 P_{ij}\,\dfrac{\partial H_k}{\partial x_i}\dfrac{\partial H_m}{\partial x_j}=0\,,\qquad k,m=1,\ldots,4,
\end{equation}
where $[.,.]$ is the Schouten bracket. Remind, that the Schouten bracket $[A,B]$ of two bivectors $A$ and $B$
is a trivector which entries in local coordinates $x$ on $\mathcal M$ are equal to
\begin{equation}\label{sh-br}
[A,B]_{ijk}=-\sum\limits_{m=1}^{dim\, \mathcal M}\left(B_{mk}\dfrac{\partial A_{ij}}{\partial x_m}
+A_{mk} \dfrac{\partial B_{ij}}{\partial x_m}+\mathrm{cycle}(i,j,k)\right).
\end{equation}
If $[P,P]=0$, bivector $P$ is a Poisson bivector.

The brute force method consists of a direct solution of the equations (\ref{geom-eq}) with respect to entries of bivector $P$ using an appropriate anzats. Remind, that  a'priory these equations have infinitely many solutions \cite{ts07} and using of anzats allows us to get some partial solutions in a constructive way.   Below we use the following simple anzats
\begin{equation} \label{p-anz}
P_{ij}=\sum_{k=1}^6 c_{ij}^k(\g)M_k+d_{ij}(\g)\,,
\end{equation}
  where $c_{ij}^k$ and $d_{ij}$ are unknown functions on $\g$.

Points on the phase space manifold where the rank of $P$ is full are called regular points, and those where the rank is less than full, singular. Usually the change in rank governs physical properties like the presence of extra equilibria, or the stability of existing equilibria. Center of the Poisson bracket algebra consists of the Casimir functions which are in the involution with all other functions.  The level sets of these Casimir functions would locally carve out symplectic leaves of even dimension equal to the rank of the Poisson bivector. When rank changes occur, these leaves drop in dimension by an even integer: extra Casimir functions, called subcasimir functions, arise and the new symplectic leaves of reduced dimension (lower dimensional strata or so-called thin orbits) are defined by the intersection of level sets of both the Casimir  and the subcasimir functions. Physical systems tend to equilibrate towards states of greater symmetry, which occur on dynamical leaves of greater codimension. Thus, singular leaves become relevant as arenas where actual stability issues of equilibria must be addressed, see \cite{nar13} and references within.

In \cite{ts08l} we successfully solved equations (\ref{geom-eq}) for the Lagrange top, which has the same symmetry vector field $X_S$ and well-known linear integrals of motion $J_1=M_3$ and $J_2=(\g,M)$. For one of these solutions  function $C=(\g,\g)$ is not the Casimir function and, therefore, we will not consider a five-dimensional submanifold defined by the physical condition $C=1$ in order to preserve generic mathematical construction, see also \cite{ts07a}.  In \cite{bts12,ts11,ts12a,ts12b} reader can also find  other solutions of the  equations (\ref{geom-eq}) for various nonholonomic systems.

\section{Poisson brackets for body of revolution on a plane}

Substituting explicit definitions of the Hamilton function $H$ and the geometric integral $C$ and implicit definitions of the remaining two integrals of motion $J_{1,2}$ (\ref{rotb-addint}) into the equations  (\ref{geom-eq}) one gets overdetermined system of algebro-differential equations on coefficients $c_{ij}^k$ and $d_{ij}$ of no more than linear polynomials in momenta $P_{ij}$ (\ref{p-anz}).

In contrast with the Lagrange top in the nonholonomic  case  the resulting equations have only two solutions
\begin{equation}\label{rotb-pg}
P^{(k)}=\alpha\, P_\alpha^{(k)}+\beta\, P_\beta\,,\qquad  k=1,2
\end{equation}
with  rank$P^{(k)}=4$ if $\g_3\neq\pm 1$. Here  $P_\alpha$ and $P_\beta$ are compatible rank-two Poisson bivectors
\[[P_\alpha,P_\beta]=0\,\]
satisfying (\ref{geom-eq}), whereas $\alpha$ and $\beta$ are arbitrary functions on $\g_2/\g_1$ and $\g_3$, respectively. It allows us to say  that  solutions (\ref{rotb-pg}) of (\ref{geom-eq}) are decomposed into "horizontal" and "vertical" parts with respect to the action of the symmetry subgroup, similar to decomposition of the vector field $X$, see \cite{bl11} and \cite{bates93,ekm04,koi92}.

First parts $ P_\alpha^{(k)}$  are labelled by pairs of functions $v_{1,2}^{(k)}$ entering into the definition of the linear integral of motion $J_{k}$ (\ref{rotb-addint})
\begin{equation}\label{vert-p}
 P_\alpha^{(k)}=\zeta_k \left(
    \begin{array}{cc}
     0 & \mathbf\Gamma_\alpha \\
    - \mathbf\Gamma^\top_\alpha & \mathbf M_\alpha \\
     \end{array}
    \right)\,,\qquad \mbox{ rank}\,P_\alpha^{(k)}=2\,,\qquad k=1,2,
\end{equation}
where
\begin{eqnarray}
\mathbf\Gamma_\alpha&=&\left(
   \begin{array}{ccc}
    \frac{\g_1\g_2}{\g_1^2+\g_2^2}\frac{v_2^{(k)}}{v_1^{(k)}} & \frac{\g_2^2}{\g_1^2+\g_2^2} \frac{v_2^{(k)}}{v_1^{(k)}} & -\scriptstyle\g_2 \\
   - \frac{\g_1^2}{\g_1^2+\g_2^2}\frac{v_2^{(k)}}{v_1^{(k)}} & -\frac{\g_1\g_2}{\g_1^2+\g_2^2}\frac{v_2^{(k)}}{v_1^{(k)}} &\scriptstyle \g_1 \\
    0& 0 & 0 \\
   \end{array}
   \right)\,,\nonumber\\
 \mathbf M_\alpha&=&\left(
   \begin{array}{ccc}
    0& \frac{\g_1M_1+\g_2M_2}{\g_1^2+\g_2^2} \frac{v_2^{(k)}}{v_1^{(k)}} & -\scriptstyle M_2 \\
   * &0 & \scriptstyle M_1 \\
   *& * & 0 \\
   \end{array}
   \right)
\nonumber
\end{eqnarray}
and
\[\begin{array}{l}\scriptscriptstyle
\zeta_k =\exp\left(\bigintss\frac{m\Bigl(
mf_1^2\bigl((1-\g_3^2)f_2+\g_3f_2\bigr)(f_1v_2^{(k)}-f_2v_1^{(k)})+I_1v_1^{(k)}f_1\bigl((1-\g_3^2)f'_1-\g_3f_1\bigr)+I_3v_2^{(k)}(f'_1f_2-f_1^2)
\Bigr)}{v_1^{(k)}g^2}\,\mathrm d\g_3\right)\,.
\end{array}
\]
These rank-two Poisson bivectors have the following Casimir functions
\[  P_\alpha^{(k)}\mathrm dC=0\,,\quad P_\alpha^{(k)}\mathrm dJ_k=0\,,\quad P_\alpha^{(k)}\mathrm d\g_3=0\,,\quad P_\alpha^{(k)}\mathrm d(\g_1M_2-\g_2M_1)=0\]
and give rise to the symmetry field acting on $M_3$
\[ P_\alpha^{(k)}\mathrm dM_3=X_S\,.\]
We can consider $P_\alpha^{(k)}$ as the "vertical" parts of  $P^{(k)}$ (\ref{rotb-pg}), which will be lost taking the quotient by  symmetry subgroup $SO(2)$.

Second part of (\ref{rotb-pg}) looks like
 \begin{equation}\label{rotb-pu}
P_\beta=\left(
     \begin{array}{cc}
     0 & \mathbf\Gamma_\beta \\
     -\mathbf\Gamma^\top_\beta & \mathbf M_\beta \\
     \end{array}
    \right)\,,\qquad \mbox{rank}\,P_\beta=2\,,
\end{equation}
where
\[\begin{array}{l}
 \mathbf\Gamma_\beta=\left(
   \begin{array}{ccc}
    \frac{\g_1\g_2\g_3}{\g_1^2+\g_2^2} & -\frac{\g_1^2\g_3}{\g_1^2+\g_2^2} & 0 \\
    \frac{\g_2^2\g_3}{\g_1^2+\g_2^2} & -\frac{\g_1\g_2\g_3}{\g_1^2+\g_2^2} & 0 \\
    -\g_2 & \g_1 & 0 \\
   \end{array}
   \right)\,,\\
   \\
 \mathbf M_\beta=\left(
   \begin{array}{ccc}
   \scriptstyle 0 &{\scriptstyle -M_3+}\frac{\g_3(\g_1M_1+\g_2M_2)}{\g_1^2+\g_2^2}+m\tau & -m\sigma\g_2 \\
    * &\scriptstyle0 & m\sigma \g_1\\
   * & * & \scriptstyle0 \\
   \end{array}
   \right)
   \nonumber
\end{array}
\]
and
\[\begin{array}{l}
\tau=\dfrac{f_2}{f_1}\sigma-\frac{f_1\Bigr(I_3f_2(\g_1M_1+\g_2M_2)-(\g_1^2+\g_2^2)I_1f_1M_3\Bigr)}{g^2}
\left(1-\left(\dfrac{f_2}{f_1}\right)'\right)\,,\\
\\
\sigma=\frac{(\g_1M_1+\g_2M_2)\Bigl(
m(r,\g)f_1^3+I_3(f_1^2+f'_1f_2)\Bigr)+M_3\Bigl(m(r,\g)f_1^2f_2+\bigl(\g_3f_1-(\g_1^2+\g_2^2)f'_1\bigr)I_1f_1\Bigr)}{g^2}\,.
\end{array}
\]
This rank-two bivector $ P_\beta$ has the following four Casimir functions
\[  P_\beta\,\mathrm dC=0\,,\qquad P_\beta\,\mathrm dJ_1=0\,,\qquad  P_\beta\,\mathrm dJ_2=0\,,\qquad P_\beta\,\mathrm d(\g_2/\g_1)=0\,.\]
It depends only on the form of the body, i.e. only on $f_{1,2}$ and it may be considered as "horizontal" part of the complete solution (\ref{rotb-pg}), which will survive in the reduction by $SO(2)$ symmetry.

In fact bivector $P_\beta$ coincides with the standard rank-two bivector $P^{(y)}$ (\ref{eqy-ham}) if we put
\[y_1=\g_3\,,\qquad y_2=\dfrac{gM_1}{\g_2}\,,\qquad y_3=H\,,\qquad y_4=J_1\,,\qquad y_5=J_2\,,\qquad y_6=(\g,\g)\,.\]
Here $g$ is defined by (\ref{rotb-mu}).

Using freedom in a choice of functions $\beta$ and $v_{1,2} $  we can get  Poisson bivectors $P^{(k)}$ with well defined limit at  $\g_{1,2}\to 0$.  The corresponding bivectors have rank two at $\gamma_3=\pm 1$ and functions  $x_3$ and $M_3$ play the role of the subcasimir functions. This change in rank indicates the possible  presence of equilibria or the stability of existing equilibria.  Of course, for an existing of equilibria we have impose certain restrictions on the angular velocity of the body, see discussion of the Rouths sphere in \cite{cus98}.

In generic case the  rank-four Poisson bivectors $P^{(k)}$ (\ref{rotb-pg}) have the following Casimir functions
\[
  P^{(k)}\mathrm dC=0\,,\qquad P^{(k)}\mathrm dJ_k=0\,,\qquad k=1,2\,.
\]
Equality of the Schouten brackets to zero is equivalent to the existence of integrals of motion $J_{1,2}$ (\ref{rotb-eqv})
\[
[P^{(k)},P^{(k)}]=0\qquad\Leftrightarrow\qquad \dot{J_k}=X(J_k)=0\,.
\]
We can say that the existence of the Casimir function is equivalent to the existence of the integral of motion of the corresponding dynamical system..

\begin{prop}
For the rigid body of revolution vector field $X$ (\ref{rotb-eqm}) is a linear combination of the Hamiltonian vector field and the symmetry field $X_S$ (\ref{vf-symm})
\begin{equation}\label{rotb-dec}
X=\beta^{-1}P^{(k)}\mathrm dH+\eta_k\,X_S\,,\qquad\eta_k=\dfrac{u_1(\g_1M_1+\g_2M_2)+u_2M_3}{g^2},
\end{equation}
where coefficients
\[
\begin{array}{l}
u_1=\frac{mf_1(1-\g_3^2)\Bigl(\beta^{-1}\alpha\zeta_k (v_1^{(k)}f_2-v_2^{(k)}f_1)+v_1^{(k)}(\g_3f_1-f_2)\Bigr)+I_3(\g_3v_1^{(k)}+\beta^{-1}\alpha\zeta_k v_2^{(k)})}{(1-\g_3^2)v_1^{(k)}}
\\
\\
u_2=\frac{mf_2\Bigl(\beta^{-1}\alpha\zeta_k (v_1^{(k)}f_2-v_2^{(k)}f_1)+v_1^{(k)}(\g_3f_1-f_2)\Bigr)-I_1v_1^{(k)}(\beta^{-1}\alpha\zeta_k +1)}{v_1^{(k)}}\,,\\
\end{array}
\]
depend on a pair of functions
 $v_{1,2}^{(k)}$ entering into $J_{k}$ (\ref{rotb-addint}).
\end{prop}
The proof is a straightforward verification of the equations (\ref{rotb-dec}).

It is easy to prove that
\[u_1=u_2=0 \qquad\Rightarrow\qquad X=\beta^{-1}P^{(k)}\mathrm dH\]
only if $v_2^{(k)}=\g_3\,v_1^{(k)}$. It means that vector field $X$ is the Hamiltonian vector field only if there is integral of motion
\[
J=v_1^{(k)}(\g,M)\,,
\]
which is proportional to the standard Casimir function $(\g,M)$ of canonical Poisson brackets on the Lie algebra $e^*(3)$.

On the other hand, if we take $\alpha=const$ and
\begin{equation}\label{rotb-beta}
\beta=-\dfrac{(1-\g_3^2)\bigl(f_1v_2^{(k)}-f_2v_1^{(k)}\bigr)^2m+I_1(1-\g_3^2){v_1^{(k)}}^2+I_3{v_2^{(k)}}^2}
{(1-\g_3^2)(\g_3f_1-f_2)\bigl(f_1v_2^{(k)}-f_2v_1^{(k)}\bigr)m+I_1(1-\g_3^2)v_1^{(k)}+I_3v_2^{(k)}}\,\dfrac{\alpha\zeta_k}{v_1^{(k)}}\,,
\end{equation}
then we get
\begin{equation}\label{conf-j}
X=\beta^{-1}P^{(k)}\mathrm dH+\delta J_kX_S\,,
\end{equation}
where
\[
\delta=\dfrac{\g_3v_1^{(k)}-v_2^{(k)}}{(1-\g_3^2)\bigl(f_1v_2^{(k)}-f_2v_1^{(k)}\bigr)^2m+I_1(1-\g_3^2){v_1^{(k)}}^2+I_3{v_2^{(k)}}^2}\,.
\]
It allows us to prove the following
\begin{prop}
For the rigid body of revolution vector field $X$ (\ref{rotb-eqm}) is the conformal Hamiltonian vector field
\[X=\beta^{-1}P^{(k)}\mathrm dH\]
 on the zero level of integrals of motion
 \[J_k=0, \qquad k=1,2\,,\]
 where $\beta$ is given by (\ref{rotb-beta}).
\end{prop}
In generic case in order to get the Hamiltonian vector field we can use the Chaplygin Hamiltonization, which is discussed in the next paragraph.

\subsection{The Chaplygin Hamiltonization}

Following to \cite{ch48,bm02} let us consider  the ultimate four-dimensional reduced phase space $\mathcal M/G$ with coordinates $\g_3$ and
\begin{equation}\label{rotb-K}
K_1=\dfrac{(M,r)}{f_1}\,,\qquad K_2=g{\omega_3}\,,\qquad K_3=\dfrac{\kappa(\g_1M_2-\g_2M_1)}{1-\g_3^2}\,,
\end{equation}
where
 \[\kappa=\sqrt{\dfrac{1-\g_3^3}{I_1+m(r,r)}}\,.\]
These coordinates are defined only at $\g_3\neq \pm 1$. Both bivectors $P^{(1,2)}$ (\ref{rotb-pg}) have a common projection on this subspace of the six-dimensional space with coordinates $x=(\g,M)$. At $\beta=1$ it looks like
\begin{equation}\label{dp-rotb}
\hat{P}=\kappa\left(
   \begin{array}{cccc}
   0 & 0 & 0& 1 \\
   0 & 0 & 0 & - I_3 g^{-1}\Bigl(1-(f_2/f_1)'\Bigr)K_2 \\ \\
   0 & 0 & 0 & - mg^{-1} f_1(f_1-f'_2)K_1 \\ \\
   * & * & * & 0 \\
   \end{array}
  \right)\,,\qquad \mbox{rank}\,\hat{P}=2\,.
\end{equation}

Projection of the initial vector field $X$ (\ref{rotb-eqm},\ref{rotb-dec}) on this subspace  becomes the Hamiltonian vector field
\[\hat{X}=\hat{P}\mathrm dH\]
with respect to this degenerate Poisson structure. Dividing two Hamiltonian equations of motion
\[
 \dot{K}_1=-\kappa g^{-1} I_3\left(1-\left(\dfrac{f_2}{f_1}\right)'\right)\,K_2K_3\,,\qquad
\dot{K_2}=-\kappa m g^{-1} f_1(f_1-f'_2)\,K_1K_3\,,
\]
by the third Hamiltonian equation of motion
\[\dot{\g}_3=\{H,\g_3\}=\kappa K_3\,,\]
one gets a system of linear non autonomous first order differential equations
\begin{equation}\label{naeq-rotb}
\dfrac{\mathrm dK_1}{\mathrm d\g_3}=-g^{-1}I_3\left(1-\left(\dfrac{f_2}{f_1}\right)'\right)K_2\,,\qquad
\dfrac{\mathrm dK_2}{\mathrm d\g_3}=-mg^{-1}f_1(f_1-f'_2)K_1\,,
\end{equation}
which was obtained by Chaplygin in \cite{ch48}.
The generic solution of this system reads as
\begin{equation}\label{inv-caz-rotb}
K_i=c_1 \Phi_1(\g_3)+c_2 \Phi_2(\g_3)\,,\qquad i=1,2,
\end{equation}
where $ \Phi_{1,2}$ are fundamental solutions of (\ref{naeq-rotb}), whereas constants $c_{1,2}$ are the values of integrals of motion $J_{2,3}$ (\ref{rotb-addint}). Solving (\ref{inv-caz-rotb}) with respect to $c_{1,2}$ we could get these integrals which are the Casimir functions of the Poisson bivector $\hat{P}$, see details in \cite{bm02}.

The Poisson bivector $\hat{P}$ was obtained in \cite{bm02} using Chaplygin's reducing multiplier theory. Similar rank-two Poisson structures are discussed in \cite{fass05,herm95,mosh87,ram04}.

\subsection{Reduction to the canonical Poisson brackets}

Let us consider the following transformation of momenta
\begin{equation}\label{rotb-map}
\begin{array}{l}
L_1=\frac{1}{\g_1^2+\g_2^2}\left(\frac{\g_1\g_3}{\alpha}\left(\dfrac{v_1^{(k)}(\g_1M_1+\g_2M_2)}{v_2^{(k)}\zeta_k}
+b\varrho_k\right)+\frac{\g_2(\g_1M_2-\g_2M_1)}{\beta}+c\g_1\right)\,,\nonumber\\
\nonumber\\
L_2=\frac{1}{\g_1^2+\g_2^2}\left(\frac{\g_2\g_3}{\alpha}\left(\dfrac{v_1^{(k)}(\g_1M_1+\g_2M_2)}{v_2^{(k)}\zeta_k}
+b\varrho_k\right)-\frac{\g_1(\g_1M_2-\g_2M_1)}{\beta}+c\g_2\right)\,,\nonumber\\
\nonumber\\
L_3=\frac{M_3}{\alpha\zeta}-\frac{1+\zeta_k\varrho_k v_2^{(k)}}{\alpha\zeta_k v_2^{(k)}}\,b\,, \qquad \qquad b=J_k\,,\quad c=(L,\g)\,,
\end{array}
\end{equation}
where
\begin{equation}\label{rotb-varrho}
\varrho_k=\int \dfrac{v_1^{(k)}\,\mu_k}{{v_2^{(k)}}^2\zeta_k}\,\mathrm d\g_3\,,
\end{equation}
and
\[
\mu_k=1-mg^{-2}(1-\g_3^2)(I_1+mf_2^2)f_1^2+mg^{-2}\Bigl(m\g_3 f_2^3-I_1(1-\g_3^2) f'_2+I_1\g_3f_2\Bigr)f_1\,.
\]

\begin{prop}
If $\alpha=const$ the Poisson map (\ref{rotb-map})
\[
\psi:\qquad (\g,M)\to (\g,L)
\]
 reduces the Poisson brackets $\{.,.\}$ associated with bivectors $P^{(k)}$ (\ref{rotb-pg}) to the canonical Poisson brackets on the Lie algebra
 $e^*(3)$
\begin{equation}\label{e3}
\bigl\{L_i\,,L_j\,\bigr\}_0=\varepsilon_{ij\ell}L_\ell\,,
 \qquad
\bigl\{L_i\,,\g_j\,\bigr\}_0=\varepsilon_{ij\ell}\g_\ell \,,
\qquad
\bigl\{\g_i\,,\g_j\,\bigr\}_0=0\,,
\end{equation}
where $\varepsilon_{ij\ell}$ is a completely antisymmetric tensor.
\end{prop}
The proof consists of substituting momenta $L_i$ (\ref{rotb-map}) into the brackets (\ref{e3}).

The  Poisson map (\ref{rotb-map}) is defined on the  regular part of the Poisson manifold  without the singular points
$\g_3=\pm1$.  However we suppose that  this singularity  isn't connected with a possible existence of equilibria because the same singularity at the Poisson map we found for the Chaplygin ball,  Veselova systems \cite{ts12b,ts12c} and generalized Chaplygin ball \cite{ts13}. It will be interesting to study such common properties of various deformations of the canonical Lie-Poisson brackets associated with different nonholonomic models.

We can use this reduction of the Poisson bracket to a canonical one in order to prove equivalence of some nonholonomic systems. For instance, in \cite{ts12b,ts12c} such reduction of the Poisson brackets associated with the Chaplygin ball and Veselova system allowed us to prove trajectory equivalence of these nonholonomic systems.

We can also use this reduction in order to solve equations of motion. Namely, it is well-known that symplectic leaf of $e^*(3)$ fixed by $(\g,\g)=1$ and $(\g,L)=0$ is symplectomorphic to the cotangent bundle $T^*S$ to two-dimensional unit sphere $S$. So, at
\[c=(\g,L)=0\] we have the Poisson map for our six-dimensional phase space to the cotangent bundle $T^*S$ with the canonical Poisson brackets. For the Poisson bivector $P^{(1)}$ with Casimir function $J_1$ (\ref{rotb-addint})
\[P^{(1)}\mathrm dJ_1=0\]
the images of the Hamilton function $H$ and second linear integral $J_2$ have the following form
\begin{eqnarray}\label{rotb-Hsph}
H&=&AL_3^2+B(L_1\g_2-L_2\g_1)^2+bCL_3+b^2D\,,\\
\nonumber\\
J_2&=&\dfrac{\zeta_1(v_1^{(1)}v_2^{(2)}-v_1^{(2)}v_2^{(1)})}{v_1^{(1)}}\,(\alpha L_3+b\varrho_1)+\frac{v_2^{(2)}}{v_2^{(1)}}\,b\,,\nonumber\\
\end{eqnarray}
where $A,B,C$ and $D$ are functions on $\g_3$.

The corresponding Hamiltonian equation of motion for $\g_3$ is solved completely similar to the Lagrange top. Because $\dot{\g_3}=\{H,\g_3\}$ and coefficients of (\ref{rotb-dec}) depend only on $\g_3$ we can solve the remaining equations of motion by quadratures too.

However,  our main aim will be identification of the common level surfaces of these integrals with bi-lagrangian foliation using a concept of the natural Poisson bivectors on the Riemannian manifolds \cite{ts11a}. After that we could to apply  inverse to (\ref{rotb-map}) transformation to the corresponding Poisson pencil and to get bi-Hamiltonian description of the initial non-holonomic model.

\subsection{Gyrostatic generalizations}
According to S.A. Chaplygin \cite{ch48} we can add to the body the uniformly rotating balanced rotor. The corresponding system can be interpreted as a nonholonomic gyrostat. The gyrostatic effect can also be obtained by an adding multiply connected cavities completely filled with the ideal incompressible liquid possessing nonzero circulation in the body.

In this case the equations of motion are equal to
\begin{equation}\label{rotb-eqm-gyr}
\dot M=(M+S)\times \omega+m\dot{r}\times(\omega\times r)+M_F\,,\qquad
\dot \g=\g\times \omega\,,
\end{equation}
where $S$ is the constant three-dimensional vector of gyrostatic moment.

For the body of revolution we suppose that gyrostatic moment
\begin{equation}\label{rotb-s} S = (0, 0, s)\end{equation}
is directed along the axis of revolution. In this case linear integrals of motion (\ref{rotb-addint}) are shifted by linear in $s$ term
\[
J^{(s)}_{k}=J_k+s\int v_1^{(k)}\mathrm d\g_3\,,\qquad k=1,2.
\]
Here $v_1^ {(k)} $ are functions on $\gamma_3$ from the definition of integrals $J_k$ (\ref{rotb-addint}) for the body without rotor.

In this case even for a disk (\ref{disk-int}) one gets non algebraic integrals of motion $J_k$ because integrals $\int L^{(1,2)}\left(b_-,\g_3\right) \mathrm d\g_3$ on the Legendre functions $L^{(1,2)}$ are non algebraic functions.

\begin{prop}
For the nonholonomic gyrostat solutions of the equations (\ref{geom-eq}) have the form
\begin{equation}\label{gyr-pg}
P^{(k)}_s=P^{(k)}-\beta\,\left(
    \begin{array}{cc}
    0 & 0 \\
    0 & \mathbf S
    \end{array}
    \right)\,
\end{equation}
where
\[
\mathbf S=\left(
    \begin{array}{ccc}
    0 & S_3 & -S_2 \\
    -S_3 & 0 & S_1 \\
    S_2 & -S_1 & 0
    \end{array}
   \right)\]
 is a $3\times 3$ skew symmetric matrix associated with three-dimensional vector $S$. In our case (\ref{rotb-s}) $S_3=s$ and $S_1=S_2=0$\,.

The corresponding vector field is a sum of the Hamiltonian vector fields and symmetry vector field
\[
X=\beta^{-1}P_s^{(k)}\mathrm dH+\eta_k\,X_S\,,\qquad k=1,2,
\]
with the same coefficients $\eta_k$ as for the body without rotor.
\end{prop}
It is interesting that for the gyrostatic generalisation of the Chaplygin ball we can use the absolutely same shift of the corresponding Poisson bivector.

\subsection{Example}\label{section-routh}

Let us consider the Routh sphere, which is one of the well known examples of the rigid body of revolution \cite{bts12,bm02,ch11,cus98,herm95,ram04}. The center of mass of this sphere is shifted with respect to its geometric center and the line joining the center of mass and the geometric center is an axis of inertial symmetry. It means that in the plane perpendicular to this axis the moment of inertia tensor has two equal principal moments of inertia. This sphere rolls on a horizontal sphere under the influence of a constant vertical gravitational force.

In this case $f_1=R$ and $f_2= R\g_3+a$, i.e.
\[ r=(R\g_1,R\g_2, R\g_3+a)\,,\]
where $R$ is a radius of the ball and $a$ is a distance from the geometric center to the center of mass.
As for the symmetry Lagrange top there are two linear in momentum integrals of motion. The first integral
\begin{equation}\label{jell-int}
J_1=(M,r)
\end{equation}
 is a well-known Jellet integral \cite{jell72}, see also \S243, p. 192 in Routh's book \cite{rou55} .
The second integral
\begin{equation}\label{raus-int}
{J}_2=g(\g)\,\omega_3\,,
\end{equation}
was found by Routh in 1884 \cite{rou55} and recovered later by Chaplygin in \cite{ch48}.

In this case we have
\[\begin{array}{lll}
v_1^{(1)}=f_1=R\,,\qquad & v_2^{(1)}=f_2=R\g_3+a\,,\qquad &\zeta_1 =g\,,\\
\\
v_1^{(2)}=\dfrac{mR(R\g_3+a)}{g}\,,\qquad & v_2^{(2)}=\dfrac{I_1+m(R\g_3+a)^2}{g}\,,\qquad &\zeta_2 =R\g_3+a\,.
\end{array}
\]
The corresponding Poisson bivectors $P^{(1,2)}$ (\ref{rotb-pg}) were found in \cite{bts12}.

These bivectors $P^{(1,2)}$ (\ref{rotb-pg}) allows us to get the following representations of the initial vector field $X$ (\ref{rotb-eqm})
\begin{equation}\label{sphere-dec}
X=\beta^{-1}P^{(1)}\mathrm dH+\eta_1X_S=\beta^{-1}P^{(2)}\mathrm dH+\eta_2X_S\,,
\end{equation}
or
\begin{equation}\label{sphere-biham}
X=\beta^{-1}P^{(1)}\mathrm dH+\dfrac{\alpha I_1}{g^2}\,\eta_1\,P^{(1)}\mathrm dJ_2\nonumber\\
=\beta^{-1}P^{(2)}-\dfrac{\alpha I_1}{mg^2}\,\eta_2\,P^{(2)}\mathrm dJ_1\,.
\end{equation}
The last equation may be considered as a nonholonomic counterpart of the standard Lenard-Magri recurrence relations
\[X=P\mathrm dH_1=f_1P'\mathrm dH_2+f_2P'\mathrm dH_3\]
for two dimensional bi-Hamiltonian systems $(f_1=1,\,\quad f_2=0)$, quasi bi-Ha\-mil\-to\-ni\-an systems $(f_2=0)$ or bi-integrable systems $(\forall f_{1,2})$, which appear in Hamiltonian mechanics \cite{ts07,ts08l,ts11}.

Let us consider the Poisson map (\ref{rotb-map}) associated with the Jellet integral of motion (\ref{jell-int}) where
\[
\varrho_1=-\frac{I_1+m(R\g_3+a)^2}{gI_1(R\g_3+a)}\,.
\]
At $\alpha=conts$ and $c=0$ the Hamilton function
\begin{eqnarray}
H&=&\frac{1}{2R^2(1-\g_3^2)}\left(
\alpha^2L_3^2\bigl(I_1R^2(\g_3^2-1)-(R\g_3+a)^2I_3\bigr)+
\frac{\beta^2(\g_2L_1-\g_1L_2)^2}{\bigl(I_1+m(r,r)\bigr)}\right.\nonumber\\
&-&\left.
\frac{2\alpha g b(r\g_3+a)L_3}{I_1}+\frac{I_1+m(R\g_3+a)^2b^2}{I_1^2}
\right)+U(\g_3)
\label{ham-jell}
\end{eqnarray}
and the Routh integral of motion
\[
J_2=\alpha I_1 L_3\,,
\]
 define an integrable system on the cotangent bundle $T^*S$. In standard
 spherical coordinates
\begin{equation}\label{sph-coord}
\begin{array}{ll}
\g_1 =\sin\phi\sin\theta,\qquad &
L_1 =\dfrac{\sin\phi\cos\theta}{\sin\theta}\,p_\phi-\cos\phi\,p_\theta\,,\\
\\
\g_2 = \cos\phi\sin\theta,\qquad&
L_2=\dfrac{\cos\phi\cos\theta}{\sin\theta}\,p_\phi+\sin\phi\,p_\theta\,, \\
\\
\g_3 =\cos\theta\,,\qquad &L_3 = -{p_\phi}\,,
\end{array}
\end{equation}
where $\phi,\theta$ are the Euler angles, $p_\phi$ and $p_\theta$ are the canonically conjugated momenta
\[
\{\phi,p_\phi\}=\{\theta,p_\theta\}=1\,,\qquad \{\phi,\theta\}=\{\phi,p_\theta\}=\{\theta,p_\phi\}=0\,,\]
 integrals of motion look like
\begin{equation}\label{int-sph2}
{H}=\frac{A(\theta)\,p_\phi^2+B(\theta)\,p_\theta^2+bC(\theta)\,p_\phi+b^2D(\theta)}{2}+U(\theta)\,,\qquad J_2=-\alpha I_1\,p_\phi\,.
\end{equation}
Here $b$ is a value of the Jellet integral $J_1$, $\alpha=const$ and
\[\begin{array}{ll}
A(\theta)=\alpha^2\left(I_1+\frac{I_3(a^2+2aR\cos\theta+R^2\cos^2\theta)}{R^2\sin^2\theta}\right)\,,\quad&
B(\theta)=\frac{\beta^2}{I_1+m(a^2+2aR\cos\theta+R^2)}\,,\\
\\
C(\theta)=\frac{2\alpha g(R\cos\theta+a)}{I_1R^2\sin^2\theta}\,,\qquad &
D(\theta)=\frac{I_1+m(a^2+2aR\cos\theta+R^2)}{I_1^2R^2\sin^2\theta}\,.
\end{array}
\]
Using the expansion of the initial vector field (\ref{sphere-dec}) one gets
\[
\dot{\theta}=\{H,\theta\}\,.
\]
Thus, similar to the Lagrange top, we have a standard equation for the nutation angle
\[
\dot{\theta}=B(\theta)\,p_\theta=\sqrt{B(\theta)\Bigl(2E_1-A(\theta)E_2^2-bE_2C(\theta)-b^2 D(\theta)-2U(\theta)\Bigr)}\,,
\]
where $E_1={H}$ and $E_2=-J_2/\alpha I_1$ are constants of motion. Solving this equation by quadrature one gets equation for the second Euler angle
\[
\dot{\phi}=\frac{(I_1\sin^2\theta+I_3\cos^2\theta)+aI_3\cos\theta}{g(\theta)I_1R\sin^2\theta} \,d -\frac{\cos\theta}{I_1R\sin^2\theta}\,b\,.
\]
Of course, we can obtain these equations without the notion of the Poisson structure and without decomposition
(\ref{sphere-dec}) of the initial vector field $X$ on the Hamiltonian and symmetric components.

Gyrostatic generalization of the Jellet integral of motion looks like
\[
J_1^s=J_1+sR\g_3\,.
\]
The Routh integral we have to shift by the non algebraic term
\[
J_2^s=J_2+s\left(
\frac{\arctan\left(\frac{\sqrt{m}(RI_1\g_3-(R\g_3-a)I_3)}{g\sqrt{I_1-I_3}}\right) a}{(I_1-I_3)^{3/2}\sqrt{m}R^2}
-\frac{g}{(I_1-I_3)mI_1R^2}
\right)\,.
\]
It is an example of the non algebraic Casimir function for the Poisson bivector $P_s^{(2)}$ (\ref{gyr-pg}) with pure algebraic entries.

%%%%%%%%%%%%%%%%%%%%%%%%%%%%%%%%%%%%%%%%%%%%%%%
\section{Poisson brackets for ball on a surface of revolution}

Like in the previous Section equations we have to substitute integrals of motion and anzats (\ref{p-anz}) into  the system of equations (\ref{geom-eq}) and to solve the resulting equations with respect to the functions $c_{ij}^k$ and $d_{ij}$ on $\g$.

The desired solutions of (\ref{geom-eq}) consist of two parts
\begin{equation}\label{ball-pg}
P^{(k)}=\alpha\, P_\alpha^{(k)}+\beta\, P_\beta\,,\qquad k=1,2,
\end{equation}
with  rank$P^{(k)}=4$ if $\g_3\neq\pm 1$. Here  $P_\alpha$ and $P_\beta$ are compatible rank-two Poisson bivectors satisfying (\ref{geom-eq}), whereas
 $\alpha$ and $\beta$ are arbitrary functions on $\g_2/\g_1$ and $\g_3$, respectively.

First parts of (\ref{ball-pg}) look like the previous bivectors  $P_\alpha^{(k)}$ (\ref{vert-p})
 \begin{equation}\label{vert-p2}
 P_\alpha^{(k)}=\zeta_k \left(
    \begin{array}{cc}
     0 & \mathbf\Gamma_\alpha \\
    - \mathbf\Gamma^\top_\alpha & \mathbf M_\alpha \\
     \end{array}
    \right)\,,\qquad \mbox{ rank}\,P_\alpha^{(k)}=2\,,\qquad k=1,2,
\end{equation}
where
\begin{eqnarray}
\mathbf\Gamma_\alpha&=&\left(
   \begin{array}{ccc}
    \frac{\g_1\g_2}{\g_1^2+\g_2^2}\frac{v_2^{(k)}}{v_1^{(k)}} & \frac{\g_2^2}{\g_1^2+\g_2^2} \frac{v_2^{(k)}}{v_1^{(k)}} & -\scriptstyle\g_2 \\
   - \frac{\g_1^2}{\g_1^2+\g_2^2}\frac{v_2^{(k)}}{v_1^{(k)}} & -\frac{\g_1\g_2}{\g_1^2+\g_2^2}\frac{v_2^{(k)}}{v_1^{(k)}} &\scriptstyle \g_1 \\
    0& 0 & 0 \\
   \end{array}
   \right)\,,\nonumber\\
 \mathbf M_\alpha&=&\left(
   \begin{array}{ccc}
    0& \frac{\g_1M_1+\g_2M_2}{\g_1^2+\g_2^2} \frac{v_2^{(k)}}{v_1^{(k)}} & -\scriptstyle M_2 \\
   * &0 & \scriptstyle M_1 \\
   *& * & 0 \\
   \end{array}
   \right)
\nonumber
\end{eqnarray}
 up to replacement of functions $v_{1,2}^{(1,2)}$ satisfying (\ref{rotb-eqv}) by functions $v_{1,2}^{(1,2)}$ satisfying (\ref{ball-eqv}) and change functions $\zeta_k$ by
\begin{equation}\label{ball-zeta}
\zeta_k =\exp\left(\bigintss
\dfrac{d f'\bigl(v_1^{(k)}(\g_3^2-1)-\g_3v_2^{(k)}\bigr)}{v_1^{(k)} f(\mu+d)}
 \,\mathrm d\g_3\right)\,.
\end{equation}
As above these rank-two Poisson bivectors have the following Casimir functions
\[  P_\alpha^{(k)}\mathrm dC=0\,,\qquad P_\alpha^{(k)}\mathrm dJ_k=0\,,\qquad P_\alpha^{(k)}\mathrm d\g_3=0\,,\qquad P_\alpha^{(k)}\mathrm d(\g_1M_2-\g_2M_1)=0\]
and generate the symmetry field acting on $M_3$
\[ P_\alpha^{(k)}\mathrm dM_3=X_S\,.\]
The difference between the Poisson bivectors $P_\alpha^{(k)}$ for the body of revolution on the plane and for the ball on a surface of revolution is related to nonholonomic reduction by different subgroups $E (2) $ and $SO (3) $, respectively.

The corresponding "horizontal" part $P_\beta$ of (\ref{ball-pg})  is equal to
\begin{equation} \label{ball-pu}
P_\beta=\left(
     \begin{array}{cc}
     0 & \mathbf\Gamma_\beta \\
     -\mathbf\Gamma^\top_\beta & \tilde{\mathbf M}_\beta \\
     \end{array}
    \right)\,,\qquad \mbox{rank}\,P_\beta=2\,.
\end{equation}
Here $\mathbf\Gamma_\beta$ is the same matrix as in (\ref{rotb-pu}) and
\[
\tilde{ \mathbf M}_\beta=\left(
   \begin{array}{ccc}
   0 &{ -M_3} +{\sigma}& \frac{d\g_2f'\bigl(M_3(\g_3^2-1)+(\g_1M_1+\g_2M_2)\g_3\bigr)} {f(\mu+d)}\\
    * &0 &-\frac{d\g_1f'\bigl(M_3(\g_3^2-1)+(\g_1M_1+\g_2M_2)\g_3\bigr)} {f(\mu+d)}\\
   * & * & 0 \\
   \end{array}
   \right)\,,
\]
where
\[
{\sigma}=\frac{(1-\g_3^2)(\mu+ \g_3^2d)f'M_3}{\g_3f(\mu+d)} +\left(\frac{(\mu+ \g_3^2d)f'}{f(\mu+d)}-\frac{\g_3}{1-\g_3^2}\right){ (\g_1M_1+\g_2M_2)}\,.
\]
This rank-two bivector $ P_\beta$ has the following four Casimir functions
\[  P_\beta\,\mathrm dC=0\,,\qquad P_\beta\,\mathrm dJ_k=0\,,\qquad P_\beta\,\mathrm d(\g_2/\g_1)=0\,,\qquad k=1,2,\]
and  coincides with the standard rank-two bivector $P^{(y)}$ (\ref{eqy-ham}) if
\[y_1=\g_3\,,\qquad y_2=\dfrac{gM_1}{\g_2}\,,\qquad y_3=H\,,\qquad y_4=J_1\,,\qquad y_5=J_2\,,\qquad y_6=(\g,\g)\,.\]
Here $g$ is defined by (\ref{ball-invm}).

Using freedom in a choice of functions $\beta$ and $v_{1,2} $  we can get  Poisson bivectors $P^{(k)}$ with well defined limit at  $\g_{1,2}\to 0$.  The corresponding bivectors have rank two at $\gamma_3=\pm 1$ and functions  $x_3$ and $M_3$ play the role of the subcasimir functions. As above this change in rank indicates only on the possibility of existence of equilibria.

Bivectors $P^{(k)}_\alpha$ and $P_\beta$ have singularities at the points $\g_3=\pm 1$. Using a freedom in the choice of the functions $\alpha$ and $\beta$ we can get well-defined bivector $P^{(k)}$ (\ref{ball-pg}) such that  rank $P^{(k)}$=2 at $\g_3=\pm1$. These points correspond to relative equilibria where the ball sits at the bottom of the surface of revolution spinning about its axis of symmetry, see \cite{fass05}. At these points $x_3$ and $M_3$ play the role of the subcasimir functions.

In generic case the  rank-four Poisson bivectors $P^{(k)}$ (\ref{ball-pg}) have the following Casimir functions
\[
  P^{(k)}\mathrm dC=0\,,\qquad P^{(k)}\mathrm dJ_k=0\,,\qquad k=1,2\,.
\]
Equality of the Schouten brackets to zero is equivalent to the existence of integrals of motion $J_{1,2}$ (\ref{ball-eqv})
\[
[P^{(k)},P^{(k)}]=0\qquad\Leftrightarrow\qquad \dot{J_k}=X(J_k)=0\,.
\]

\begin{prop}
For the ball on a surface of revolution vector field $X$ (\ref{ball-eqm}) is a linear combination of the Hamiltonian vector field and the symmetry field $X_S$ (\ref{vf-symm})
\begin{equation}\label{ball-dec}
X=-\frac{\g_3R}{\beta\bigl(\g_3f-(1-\g_3^2)f'\bigr)}P^{(k)}\mathrm dH+{\eta}_k\,X_S\,,\qquad
{\eta}_k=\dfrac{ u_1(\g_1M_1+\g_2M_2)+ u_2M_3}{f(d+\mu)}\,,
\end{equation}
where
\[
\begin{array}{l}
{u}_1=-\dfrac{\g_3R}{1-\g_3^2}+\dfrac{\alpha\zeta_k\g_3Rf\bigl( d\g_3(1-\g_3^2)v_1^{(k)}+( \g_3^2d-\mu-d)v_2^{(k)}\bigr)}{\beta\mu v_1^{(k)}(1-\g_3^2)\bigl(\g_3f-(1-\g_3^2)f'\bigr)}\\
\\
{u}_2=R+\dfrac{\alpha\zeta_k\g_3Rf\bigl(v_1^{(k)}(\mu+ \g_3^2d)- d\g_3v_2^{(k)}\bigr)}
{\beta\mu v_1^{(k)}\bigl(\g_3f-(1-\g_3^2)f'\bigr)}\,.
\end{array}
\]
\end{prop}
The proof is a straightforward verification of the equations (\ref{ball-dec}).

Like in the previous section at $\alpha=const$ and
\begin{equation}\label{ball-beta}
\beta=-\dfrac{(1-\g_3^2)(\g_3v_1^{(k)}-v_2^{(k)})^2d
+\mu\left({v_1^{(k)}}^2+(1-\g_3^2){v_2^{(k)}}^2\right)}{\bigl(\g_3f-(1-\g_3^2)f'\bigr)
\bigl(v_1^{(k)}(1-\g_3^2)+\g_3v_2^{(k)}\bigr)}\,\dfrac{\g_3 f \alpha\zeta_k}{ \mu v_1^{(k)} }
\end{equation}
 one gets
\begin{equation}\label{conf-jb}
X=-\frac{\g_3R}{\beta\bigl(\g_3f-(1-\g_3^2)f'\bigr)}P^{(k)}\mathrm dH+\delta J_kX_S\,,
\end{equation}
where
\[
\delta=\frac{R(\g_3v_1^{(k)}-v_2^{(k)})}{(1-\g_3^2)\left(\g_3v_1^{(k)}-v_2^{(k)}\right)^2d
+\mu\left({v_1^{(k)}}^2+(1-\g_3^2){v_2^{(k)}}^2\right)}\,.
\]
It allows us to prove the following
\begin{prop}
For the ball on a surface of revolution vector field $X$ (\ref{ball-eqm}) is the conformal Hamiltonian vector field
\[X=-\frac{\g_3R}{\beta\bigl(\g_3f-(1-\g_3^2)f'\bigr)}P^{(k)}\mathrm dH\]
 on the zero level of integrals of motion
 \[J_k=0, \qquad k=1,2\,,\]
 where $\beta$ is given by (\ref{ball-beta}).
\end{prop}

\subsection{The Chaplygin Hamiltonization}
Following to \cite{bmk02} let us consider  the ultimate four-dimensional reduced phase space $\mathcal M/G$ with coordinates  $\g_3$ and
\[
K_1=f(\g,M)\,,\qquad K_2=\mu\omega_3=\frac{\mu M_3+d\g_3(\g,M)}{\mu+d}\,,\qquad K_3=\frac{\g_2M_1-\g_2M_1}{\sqrt{1-\g_3^2}}\,.
\]
These coordinates are defined only at $\g_3\neq \pm 1$.  As above both bivectors $P^{(1,2)}$ (\ref{ball-pg}) have a common projection on this subspace of the six-dimensional space with coordinates $(\g,M)$. If
\[\beta=-\frac{R}{f-\frac{1-\g_3^2}{\g_3}f'}\qquad\mbox{and}\qquad
\varkappa=\frac{\sqrt{1-\g_3^2}\,R}{f-\frac{1-\g_3^2}{\g_3}f'}
\]
this projection is equal to
\begin{equation}\label{dp-ball}
\hat{P}=\varkappa\left(
   \begin{array}{cccc}
   0 & 0 & 0& 1 \\ \\
   0 & 0 & 0 &\dfrac{f'}{\g_3}\,K_2 \\ \\
   0 & 0 & 0 &\dfrac{d}{(\mu+d)f}\,K_1 \\ \\
   * & * & * & 0 \\
   \end{array}
  \right)\,,\qquad \mbox{rank}\,\hat{P}=2\,,
\end{equation}
It is the Poisson bivector having two Casimir functions $J_{1,2}$ (\ref{ball-addint}) that allows us to rewrite the projection of the initial vector field $X$ (\ref{ball-dec}) on this subspace in the Hamiltonian form
\[\hat{X}=\hat{P}\mathrm dH\,.\]
 Dividing two Hamiltonian equations of motion
\[
 \dot{K}_1=\{H,K_1\}=-\frac{\varkappa f'}{\g_3(\mu+d)}\,K_2K_3\,,\qquad
\dot{K_2}=\{H,K_2\}=-\frac{\varkappa d}{f(\mu+d)^2} \,K_1K_3\,
\]
by the third Hamiltonian equation of motion
\[\dot{\g}_3=\{H,g_3\}=-\dfrac{\varkappa K_3}{\mu+d}\,,\]
one gets a system of linear non autonomous first order differential equations
\begin{equation}\label{naeq-ball}
\dfrac{\mathrm dK_1}{\mathrm d\g_3}=\frac{f'}{\g_3}\,K_2\,,\qquad\qquad
\dfrac{\mathrm dK_2}{\mathrm d\g_3}=\frac{d}{f(\mu+d)}K_1\,,
\end{equation}
which was obtained by Routh \cite{rou55} in other variables associated with the so-called semifixed frame, see discussion in \cite{bmk02}.
This system of linear equations always possesses two integrals which are linear in $K_{1,2}$ and proportional to $J_{1,2}$ (\ref{ball-addint}).

The Poisson bivector $\hat{P}$ was obtained in \cite{bmk02} using Chaplygin's reducing multiplier theory, see also \cite{herm95}.

\subsection{Reduction to canonical Poisson brackets}
 Like in the previous Section we can study the reduction of the Poisson bivectors $P^{(k)}$ (\ref{ball-pg}) to the canonical Poisson bivector on the Lie algebra $e^*(3)$.

Namely, let us consider the Poisson map
\[
\psi:\qquad (\g,M)\to(\g,L)
\]
defined by (\ref{rotb-map}), where
\begin{equation}\label{ball}
\varrho_k=\int\frac{v_1^{(k)}}{{v_2^{(k)}}^2\zeta_k}\Bigl(1-\dfrac{(1-\g_3^2)(\mu+\g_3^2d)f'}{\g_3f(\mu+d)}\Bigr)\mathrm d\g_3\,,
\end{equation}
function $\zeta_k$ is given by (\ref{ball-zeta}) and functions $v_{1,2}^{(1,2)}$ satisfy (\ref{ball-eqv}).  It is only defined on the  regular part of the Poisson manifold where rank $P^{(k)}=4$.

\begin{prop}
 The mapping $\psi$ reduces the Poisson structures (\ref{ball-pg}) associated with the ball on a surface of revolution to the canonical Poisson structure (\ref{e3}) on the Lie algebra $e^*(3)$ at $\alpha=const$.
\end{prop}

As above using this Poisson map we can identify initial dynamical system (\ref {ball-eqm}) with a dynamical system on the cotangent bundle $T^*S$ to sphere possessing the Hamilton function
\begin{equation}\label{ball-Hsph}
H=AL_3^2+B(L_1\g_2-L_2\g_1)^2+bCL_3+b^2D\,,
\end{equation}
and linear in momenta $L_3$ integral of motion $J_1$ or $J_2$.

Using the freedom in choosing the function $\beta$ we can always identify the coefficients $B$ in the Hamiltonians (\ref{rotb-Hsph}) and (\ref{ball-Hsph}). The question about identify of the Hamiltonian (\ref{ball-Hsph}) for the ball on some surfaces of revolution with Hamiltonian (\ref{rotb-Hsph}) for the special body of revolution on a plane up to canonical transformations of variables is still open.

\subsection{Example}
Let us suppose that the ball's center of mass is moving on the paraboloid
of revolution $z = c(x^2 + y^2)$. In this case
\[
f(\g_3)=-\frac{1}{2 c \g_3}
\]
and equations (\ref{ball-eqv}) have the following two independent solutions
\begin{equation}\label{ball-parv}
v_1^{(1,2)}=\g_3^{\pm \nu}\,,\qquad\mbox{and}\qquad v_2^{(1,2)}=\g_3^{1\pm\nu}\pm\frac{\nu}{1\pm\nu}\g_3^{\pm\nu-1}\,,
\end{equation}
where $\nu$ is a real number
\[
\nu=\sqrt{\frac{\mu}{\mu+d}}\,.
\]
It is easy to see that $H_2=J_1J_2$ is an algebraic integral of motion.

In this case we have
\[
\zeta_{1,2}=\g_3^{1\pm \nu}\qquad\mbox{and}\qquad \varrho_{1,2}=\pm\frac{(1\mp \nu)^2}{2\nu}\frac{\g_3^2\mp\nu(1-\g_3^2)}{\g_3^2\pm\nu(1-\g_3^2)}\g_3^{\mp2\nu}\,.
\]
At $c=0$ after the Poisson map (\ref{rotb-map}) associated with the first bivector $P^{(1)}$ and Casimir function $J_1=b$ one gets the Hamilton function
\[\begin{array}{l}
H=\frac{1-\nu^2}{2d(1-\g_3^2)}\left(\frac{\alpha^2\g_3^{2\nu}L_3^2}{(1-\nu)^2}+\beta^2(\g_2L_1-\g_1L_2)^2+
\frac{\alpha(2\g_3^2-1) bL_3}{\nu}+\frac{(1-\nu)^2b^2\g_3^{-2\nu}}{4\nu^2}\right)
\end{array}
\]
and the linear integral of motion
\[
J_2=\frac{2\alpha\nu}{\nu^2-1}L_3\,\nonumber
\]
on the cotangent bundle to sphere $T^*S$. Using spherical coordinates (\ref{sph-coord}) we can integrate the corresponding equations of motion on $T^*S$ by quadratures in a standard way.

This work was partially supported by RFBR grant 13-01-00061.

\end{document}